\documentclass[twocolumn,prl,aps,amsmath,amssymb,color,superscriptaddress]{revtex4}
\usepackage{stmaryrd}
\usepackage{amsmath}
\usepackage{amssymb}
\usepackage{graphicx}
\usepackage{dcolumn}
\usepackage{bm}
\usepackage{tabularx}
\usepackage{multirow}
\usepackage{color}

\begin{document}

\begin{titlepage}

\title{Two-dimensional ferromagnetic-ferroelectric multiferroics in violation of the $d^0$ rule}

\author{Hengxin Tan}
\affiliation{State Key Laboratory of Low-Dimensional Quantum Physics and Collaborative Innovation Center of Quantum Matter, Department of Physics, Tsinghua University, Beijing 100084, China}
\author{Menglei Li}
\email{limenglei@cnu.edu.cn}
\affiliation{Department of Physics, Capital Normal University, Beijing 100084, China}
\author{Haitao Liu}
\affiliation{Institute of Applied Physics and Computational Mathematics, PO Box 8009, Beijing 100088, China}
\author{Zhirong Liu}
\email{LiuZhiRong@pku.edu.cn}
\affiliation{Center for Nanochemistry, College of Chemistry and Molecular Engineering, Peking University, Beijing 100871, China}
\author{Yuanchang Li}
\email{yuancli@bit.edu.cn}
\affiliation{Advanced Research Institute of Multidisciplinary Science, Beijing Institute of Technology, Beijing 100081, China}
\author{Wenhui Duan}
\affiliation{State Key Laboratory of Low-Dimensional Quantum Physics and Collaborative Innovation Center of Quantum Matter, Department of Physics, Tsinghua University, Beijing 100084, China}
\affiliation{Institute for Advanced Study, Tsinghua University, Beijing 100084, China}

\begin{abstract}
Contribution of $d$-electron to ferroelectricity of type-II multiferroics causes strong magneto-electric coupling and distinguishes them from the conventional type-I multiferroics.
However, their therein polarization is too small because the ferroelectricity is merely a derivative from the magnetic order.
Here we report a new class of multiferroic materials, monolayer VO$X_2$ ($X$ = Cl, Br, and I),
which combine the advantages of type-I and type-II multiferroics.
Both ferroelectricity and magnetism arise from the same V cation,
where the filled $d$-orbital is perpendicular to an \emph{a priori} ferroelectric polarization and
thus poses no hindrance to ferroelectricity, indicating a violation of the usual $d^0$ rule.
This makes the combination of large polarizations and strong magneto-electric coupling possible.
Our findings not only add new ferromagnetic-ferroelectric multiferroics, but also point to a unique mechanism to engineer multiferroics.
\end{abstract}
\maketitle

\draft

\vspace{2mm}

\end{titlepage}
Multiferroics possessing both magnetic and ferroelectric orders\cite{Ferro162p317,Nature442p759}
are of great importance because interactions between the magnetic and electric polarizations lead to
multifarious physical effects and potential applications.
Of particular interest is the ferromagnetic-ferroelectric case for new device paradigm based on four logic states \cite{Science309p391,NatMater6p296,NatMater6p256}.
Unfortunately, ferromagnetic-ferroelectric multiferroics are rare in nature and their design was proved unexpectedly difficult \cite{Science309p391}.
In recent years, the boom of multiferroic research has been reignited,
one key factor of which is the discovery of fascinating properties in thin film systems \cite{RMP77p1083,AM23p1062,NRM2p16087}, e.g., enlarged polarizations \cite{Science299p1719,Science306p1005,NatMater6p21,AdvPhy64p519}. The search for multiferroics in the two-dimensional (2D) limit is further stimulated after the observation of stable ferroelectricity with enhanced transition temperature in monolayer SnTe \cite{Science353p274}, and robust ferromagnetism in ultra-thin transition-metal compounds of Cr$_2$Ge$_2$Te$_6$ \cite{Nature546p265} and CrI$_3$ \cite{Nature546p270}. However, up to now, only few candidates of 2D multiferroics were proposed, including magnetic-ferroelectric Hf$_2$VC$_2$F$_2$ \cite{JACS140p9768}, ferromagnetic-antiferroelectric monolayer transition-metal phosphorus chalcogenides \cite{APL113p043102} and ferromagnetic-ferroelectric electron-doped CrBr$_3$ \cite{PRL120p147601}.

Generally, the magnetism originates from the partially occupied $d$-orbitals of transition-metal cations, which, however, is believed to suppress the occurrence of ferroelectricity. This is well known as the $d^0$ rule in multiferroics \cite{JPCB104p6694}.
In this regard, the ferroelectric mechanism lies at the center of the multiferroic study.
Several alternative ferroelectric mechanisms have been reported \cite{NRM1p16046} to circumvent this contraindication, such as the 6$s$ lone-pair activity in BiFeO$_3$ \cite{Science299p1719} and PbVO$_3$ \cite{CM16p3267}, the charge ordering in the mixed valency systems of LuFe$_2$O$_4$ \cite{Nature436p1136} and Fe$_3$O$_4$ \cite{AM21p4452}, the geometric distortion in YMnO$_3$ \cite{NatMat3p164} and BaMnF$_4$ \cite{SR5p18392}, and the spin driven symmetry breaking in TbMnO$_3$ \cite{Nature426p55,PRL101p037209}.
The first three mechanisms give rise to the so-called type-I multiferroics \cite{Physics2p20} where ferroelectricity and magnetism have independent origins, resulting in large polarizations and high transition temperatures but very weak magneto-electric coupling. The last one corresponds to the type-II multiferroics \cite{Physics2p20} where the ferroelectricity is a derivative of magnetism, resulting in small polarizations and low transition temperatures despite of the favorable strong magneto-electric coupling. Ideal multiferroics should combine the advantages of the type-I and type-II multiferroics. This hightlights the uniqueness of the single-phase one-cation multiferroics where the cation would be undoubtedly responsible for both ferroelectricity and magnetism. Occurrence of such multiferroics of type-I is particularly unusual because it means some yet unrevealed interplay beyond the $d^0$ rule between $d$-electrons and ferroelectricity. Whereas, such occurrence is somewhat trivial in the type-II multiferroics because the ferroelectric polarization is only a consequence of the magnetic order, still compatible with the $d^0$ rule. As a matter of fact, only copper oxides/oxyhalides \cite{NatMater7p291,SciAdv2pe1600353,PRB82p064424,AM24p2469} are reported so far to be single-phase one-cation multiferroics, which, however, all belong to the type-II class.

In this work, based on the first-principles calculations, we identify three single-phase one-cation type-I multiferroic monolayers, including the ferromagnetic-ferroelectric VOI$_2$ and antiferromagnetic-ferroelectric VOCl$_2$ and VOBr$_2$.
Most prominently, ferroelectric properties are directly related to the V$^{4+}$ cation although it possesses a $d_{xy}$ electron, which thus violates the conventional $d^0$ rule. The origin of the coexistence rather than mutual suppression of two ferroic orders is that the occupied $d_{xy}$-orbital locates in a plane perpendicular to an \emph{a priori} polarization and thus hardly suppresses the ferroelectricity. Such characteristics make it possible to achieve the advantages of type I and type II simultaneously, i.e., large polarizations and strong magneto-electric coupling. It also points to new multiferroic physics beyond the $d^0$ rule, as well as new strategies of designing novel multiferroic materials. Note that bulk vanadium oxide dihalides with layered structures have been studied in experiments previously \cite{ZaaC479p32,JAC246p70}. The nature of van der Waals interlayer interaction makes the synthesis of 2D monolayer highly probable.

All electronic and magnetic properties are calculated within the framework of density functional theory as implemented in the Vienna $Ab$-$initio$ Simulation Package (\textsc{vasp}) \cite{VASP,PRB54p11169} with the Heyd-Scuseria-Ernzerhof (HSE) hybrid functional \cite{JCP118p8207} based on the geometric structures optimized with the Perdew-Burke-Ernzerhof (PBE) \cite{PRL77p3865} functional. Noteworthy, the HSE approach has been successfully applied in describing different metal/insulator phases of strongly correlated vanadium oxides such as VO$_2$ \cite{PRL107p016401} and V$_2$O$_3$ \cite{JCP140p054702}. The electron-ion interaction is mimicked by the projected augmented wave method \cite{PRB50p17953} with an energy cutoff of 600 eV. The $k$-point grids of 10$\times$10$\times$1 and 6$\times$6$\times$1 are employed to sample the Brillouin zone for the unit cell and 2$\times$2$\times$1 supercell, respectively. A criterion of 0.005 eV/\AA~ is used for the Hellman-Feynman forces during the structural relaxation. A vacuum layer of at least 15 \AA~ is added to avoid the spurious interaction between monolayer and its periodic images. The polarization is calculated by the berry phase method \cite{PRB47p1651}.

\begin{figure}[htb!]
 \centering
 \includegraphics[width=0.99\columnwidth]{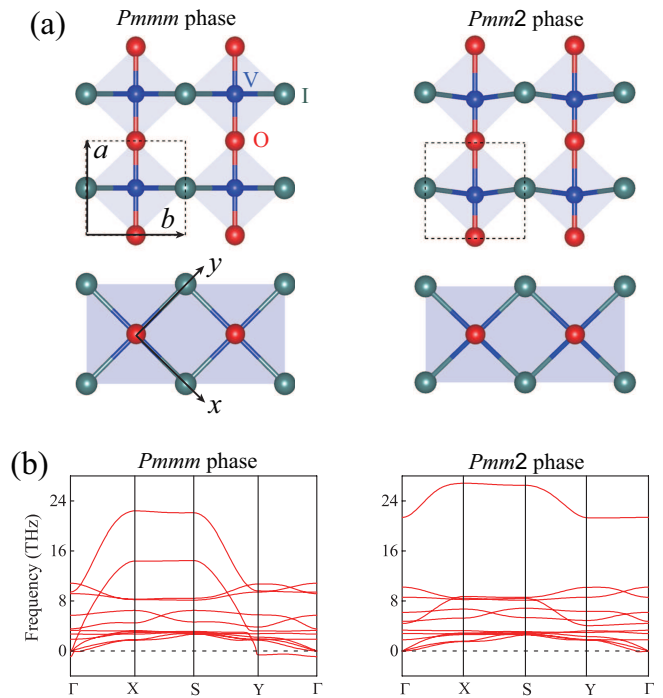}
 \caption{\label{Figure1}(Color online) Structures and phonon spectra of the layered VOI$_2$. (a) Top and front views of the high-symmetry ($Pmmm$) and low-symmetry ($Pmm2$) monolayers. Black dashed rectangle denotes the structural unit cell (containing an octahedron) without consideration of magnetic order. Blue, red and dark cyan spheres represent V, O and I atoms, respectively. The $x$ and $y$ directions are along two equivalent V-I bonds in the $Pmmm$ phase, and the $z$ directions is along the V-O bond, which will be used to illustrate the long-range magnetic order later. (b) Phonon spectra of VOI$_2$ in the $Pmmm$ and $Pmm2$ phases, respectively.}
\end{figure}

We start with the orthorhombic monolayer VO$X_2$ ($X$ = Cl, Br, and I) having inversion symmetry ($Pmmm$ space group), where two V-O (four V-$X$) bonds are of equal length with an O-V-$X$ bond angle of 90$^\circ$, as shown in Fig. \ref{Figure1}(a) for VOI$_2$ and the Supplemental Material \cite{SM} for VOCl$_2$ and VOBr$_2$. However, this paraelectric phase is unstable at zero temperature as manifested by the existence of imaginary frequency in the phonon spectrum [See Fig. \ref{Figure1}(b)], indicating a tendency of spontaneous structural distortion. The imaginary band along the $\Gamma$-Y is a polar mode corresponding to the relative movement of atoms along the V-O chain. This consequently breaks the inversion symmetry and leads to a ferroelectric polarization.

As it turns out, the distorted structure with lower $Pmm2$ symmetry removes the imaginary phonon mode [See the right panels of Fig. \ref{Figure1}]. Now two V-O bonds becomes unequal, e.g., 2.14 versus 1.67 \AA~ for the VOI$_2$. It is worth noting that the off-center displacement of V in bulk VOCl$_2$ has been observed by scanning tunnel microscope and atomic force microscope \cite{JAC246p70}. Although four V-$X$ bonds remain equivalent, the O-V-$X$ bond angle deviates from 90$^\circ$, resulting in the separation of positive and negative charge centers. This leads to polarizations of 2.8, 2.6 and 2.3 $\times$10$^{-10}$ C/m for VOCl$_2$, VOBr$_2$ and VOI$_2$ respectively [See Fig. S2 (a) in the Supplemental Material \cite{SM} for the polarization along the switching path of VOCl$_2$ as an example], which are comparable to that of the group IV monochalcogenides \cite{PRL117p097601}. Note that the polarization weakly depends on the magnetic structures which will be studied later, and the ones corresponding to the ground state are 3.0, 2.7 and 2.3 $\times$10$^{-10}$ C/m for VOCl$_2$, VOBr$_2$ and VOI$_2$, as summarized in Table I. Given their monolayer thickness of 3.3, 3.6 and 3.8 \AA, the equivalent bulk polarizations are 91, 75 and 61 $\mu$C/cm$^2$ which are at the same level as those of the traditional perovskite ferroelectrics of PbTiO$_3$ and BiFeO$_3$ \cite{NatChem8p831}, and an order of magnitude larger than those of improper ferroelectrics \cite{NatMater6p13} of the hexagonal rare-earth ferrites and manganites \cite{PRB89p205122,JPCM28p126002}.
Symmetry-lowering leads to an energy decreasing of 207, 177 and 105 meV per cation for Cl, Br and I systems. The total energy along the polarization switching path has the shape of a double well potential which is the character of ferroelectrics, as shown in Fig. S2(b) for VOCl$_2$ as an example. Such depths of the double well potential are similar to those of typical ferroelectrics such as PbTiO$_3$, BiFeO$_3$ and YMnO$_3$ \cite{PRB77p165107,PRB96p035143}, strongly implying the high stability of the ferroelectric phase.
We also calculated the intrinsic hysteresis loop of VOCl$_2$ [Fig. S2 (c)] by the approach proposed in Ref. \onlinecite{PRB66p104108}, where the determined ferroelectric coercive field is comparable to that of BaTiO$_3$.

\begin{table}
\centering
\caption{The lattice constants $a$, $b$ (\AA) and spontaneous polarization $P$ (10$^{-10}$ C/m) of the VO$X_2$ under the ground state magnetic configuration (GMC). Note that the polarizations are slightly sensitive to the long-range magnetic order, as can be seen in Table S1 of the Supplemental Material \cite{SM}.}
\renewcommand\arraystretch{1.0}
\begin{ruledtabular}
\begin{tabular}{lcccccccccccccccccccccccccc}
        & GMC & $a$ & $b$ &  $P$ \\
 \hline
   VOCl$_2$ &AFM3 & 3.815 & 3.380 & 3.0 \\
   VOBr$_2$ &AFM1 & 3.799 & 3.585 & 2.7 \\
   VOI$_2$  &FM   & 3.810 & 3.956 & 2.3 \\
\end{tabular}
\end{ruledtabular}
\begin{flushleft}
\end{flushleft}
\end{table}

\begin{figure}[htb!]
 \centering
 \includegraphics[width=0.99\columnwidth]{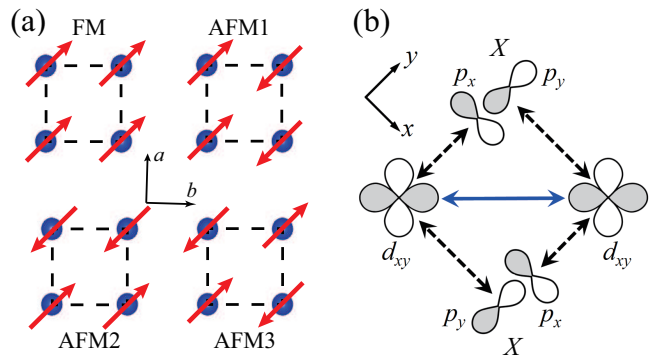}
 \caption{\label{Figure2}(Color online) (a) Four considered magnetic configurations, including the ferromagnetic (FM) and three kinds of antiferromagnetic (AFM) ones in terms of the anisotropy along the $a$ and $b$ directions. Only the V sites are shown in the 2$\times$2$\times$1 super-cell. (b) Schematic illustration of the interplay between the direct $d$-$d$ exchange (solid arrow) and two sets of halogen mediated nearly 90$^\circ$ super-exchange (dashed arrows) which determines the long-range magnetic order along $b$.}
\end{figure}

All three systems are spin-polarized with a total magnetic moment of 1 $\mu_B$, mainly coming from the V cation. We considered four magnetic configurations as illustrated in Fig. \ref{Figure2}(a) to explore the long-range magnetic order. The results are summarized in Table I (More details are found in Table S1 of the Supplemental Material \cite{SM}). A ferromagnetic ground state is achieved for the VOI$_2$, while different antiferromagnetic ones, namely, checkerboard (AFM3) and stripe (AFM1), are found for the VOCl$_2$ and VOBr$_2$, respectively. Significantly, VOI$_2$ is the only 2D material that possesses inherent ferromagnetism and ferroelectricity, so far as we know.

\begin{figure}[htb!]
 \centering
 \includegraphics[width=0.95\columnwidth]{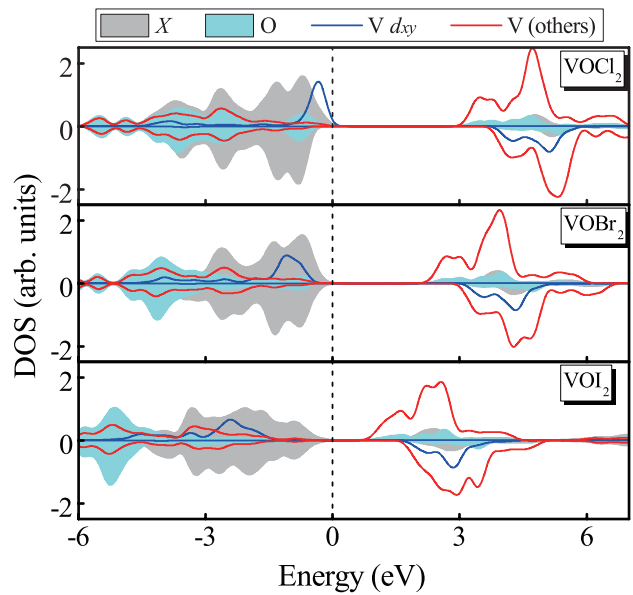}
 \caption{\label{Figure22}(Color online) Projected density of states (PDOS) of the three systems calculated using the unit cell. Note that the red lines in each panel stand for the DOS of all other V orbitals except for $d_{xy}$. The valence band maximum is set as energy zero.}
\end{figure}

Actually, these three systems are so alike except for the radii of different halogen atoms. This is reflected by the large difference between lattice constant $b$ as well as similar $a$ (See Table I). To explain their different magnetic orders, we schematically illustrate the exchange paths along  V-$X$ chains in Fig. \ref{Figure2}(b). There are two kinds of exchange interactions. One is a direct interaction between the local moments on adjacent V which favors an anti-parallel spin alignment. The other is the halogen-mediated super-exchange which favors a parallel spin alignment in terms of the Goodenough-Kanamori rule \cite{PR100p564,JPCS6p287,JPCS10p87}. Their competition determines the magnetic order. The larger the halogen radius, the larger the V-V separation along the $b$ direction, leading to the weaker direct exchange. In contrast, the electronic structures shown in Fig. \ref{Figure22} reveal an increased $p$-$d$ hybridization between V and halogen with larger radius, which implies a strengthened super-exchange from the VOCl$_2$ to VOI$_2$. For VOCl$_2$ and VOBr$_2$, the direct exchange dominates owing to the relatively small halogen radius, leading to a favourable antiferromagnetic order along the V-$X$ chains. When the direct exchange decreases to be less significant than the super-exchange, a ferromagnetic coupling becomes energetically stable, corresponding to the case of VOI$_2$.

We now turn to the central observation of this paper to examine how the ferroelectricity survives in such a non-$d^0$ system.
To reveal the role played by the very $d$-electron in VO$X_2$ where V has a $d^1$ configuration due to its +4 valency,
we first consider an analogue satisfying the $d^0$ requirement, TiO$X_2$. Similar spontaneous symmetry-lowering is observed in TiO$X_2$ to result in ferroelectricity. An example is shown in Fig. 4(a) for TiOI$_2$. Detailed geometries, phonon spectra and electronic structures of TiO$X_2$ are provided in the Supplemental Material \cite{SM}. Due to the $d^0$ nature of the Ti systems, the origin of their ferroelectricity is the same as that of the traditional perovskite BaTiO$_3$ \cite{AC5p739,PRB52p6301}, i.e., an effect driven by the phonon mode softening associated with the covalent bonding between Ti and anions. Off-center displacements of Ti endow TiOCl$_2$, TiOBr$_2$ and TiOI$_2$ with polarizations of 2.3, 2.1 and 1.9 $\times$10$^{-10}$ C/m, which are 0.7, 0.6 and 0.4 $\times$10$^{-10}$ C/m smaller than their V counterparts (values for ground states). The difference between V and Ti lies at an extra $d$-electron in V. In this regard, the occupation of $d$-orbital in VO$X_2$ enhances the ferroelectric polarization instead of inhibiting it, surprisingly opposed to the $d^0$ rule.

\begin{figure}[htb!]
 \centering
 \includegraphics[width=0.95\columnwidth]{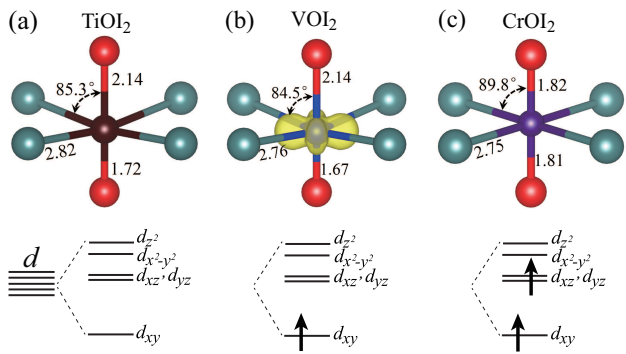}
 \caption{\label{Figure3}(Color online) Upper panels: Local geometries of the (a) TiOI$_2$, (b) VOI$_2$ and (c) CrOI$_2$. The bond lengths (in \AA) and O-metal-I bond angles are denoted. Lower panels: The corresponding $d$-orbital splitting and occupation, where an arrow means an electron. In (b), the decomposed charge density at $\Gamma$ point of the occupied $d_{xy}$-orbital (See Fig.3) is plotted in yellow with an isosurface of 0.018 $e$/Bohr$^3$.}
\end{figure}

It is the configuration of the occupied $d$-orbital that plays a central role for the survival of the polarization in VO$X_2$.
Under the local octahedral field established by oxygen and halogen, the cation $d$-orbitals split into four subgroups as revealed by our calculations: the $d_{xy}$ singlet, the $d_{xz}$ and $d_{yz}$ doublet, the $d_{x^2-y^2}$ singlet and the $d_{z^2}$ singlet. This is schematically shown in the lower panels of Fig. \ref{Figure3}. In TiO$X_2$, no $d$-orbital is occupied, so $d$-orbital does not cause any hindrance to the off-center displacements of Ti, in consistence with the $d^0$ rule. In VO$X_2$, on the other hand, the lowest $d_{xy}$ singlet is occupied. However, $d_{xy}$ is distributed in the plane perpendicular to the V-O chain, as shown by the corresponding charge density plot in Fig. \ref{Figure3}(b). Within the Slater-Koster approximation, the coupling between $d_{xy}$ and the O $p$ orbitals is zero \cite{PR94p1498,PRB92p201403}. Therefore, $d_{xy}$ does not cause any hindrance when V moves along the V-O chain. In contrast, $d_{xz}$ and $d_{yz}$ have nonzero coupling with the O $p$ orbitals \cite{PR94p1498,PRB92p201403}, so one may expect them to cause some hindrance to the polarization if occupied. As a result, the absolute energy (relative to the vacuum level) of $d_{xy}$ is less sensitive to the movement of V along the V-O chain than that of $d_{xz}$/$d_{yz}$. Indeed, the projected density of states show an almost unchanged $d_{xy}$ in the ferroelectric and paraelectric phases for VOI$_2$, and the energy shift of $d_{xy}$ is smaller than that of $d_{xz}$/$d_{yz}$ for VOCl$_2$ and VOBr$_2$ (see Fig. S4 of the Supplemental Material \cite{SM}). If one more $d$-electron is added, occupation of $d_{xz}/d_{yz}$ will occur, thus causing the suppression of polarization. This is confirmed in the analogous CrOI$_2$ monolayer where the off-center displacement of cation is almost completely removed,
as shown by the O-Cr-I bond angle close to 90$^\circ$ in Fig. \ref{Figure3}(c). Interestingly, CrOI$_2$ is a half metal (see Fig. S8 of the Supplemental Material \cite{SM} for the band structures).

In fact, the $d_{xy}$ electron also directly contributes to the polarization of VO$X_2$. This is readily understood provided that we equivalently consider VO$X_2$ as TiO$X_2$ plus an extra $d_{xy}$ electron. Although $d_{xy}$ does not couple with the O $p$ orbitals, it couples with the halogen $p$ orbitals. Owing to the existence of an \emph{a priori} ferroelectric distortion like TiO$X_2$, the hybridization between $d_{xy}$ and $X$ $p$ would cause further separation of the positive V nucleus and the negative $d_{xy}$ orbital, therefore contributing to an extra polarization. This can be quantitatively evaluated by estimating the average effective charge $\bar{q}$ of cation based on the relation, $Pab=\bar{q}\Delta x$, where $P$, $a$ and $b$ are the polarization and lattice constants of a unit cell and $\Delta x$ is the shift distance of cation from the inversion center. $\bar{q}$ of V in VOI$_2$ is determined to be 9.2$e$, while $\bar{q}$ of Ti in TiOI$_2$ is 8.6$e$. Therefore, the filled $d_{xy}$ orbital contributes about 0.6$e$ to $\bar{q}$. We also calculate their average Born effective charges and the results are 9.5$e$ and 8.7$e$, respectively, for V and Ti (see Fig. S7 of the Supplemental Material \cite{SM} for details). Difference of 0.8$e$ agrees well with the estimation, confirming the $d_{xy}$ contribution.
It is worth noting that the $d$-electron alone should not cause ferroelectric polarization, but it can enhance the ferroelectricity when parasitizing on an \emph{a priori} distortion. This opens a new route to engineer multiferroicity via the controlled magnetic doping in an intrinsic ferroelectric material.

Two characteristics distinguish VO$X_2$ from other multiferroics. On the one hand, the existence of conventional polarization endows it with the feature of the type-I multiferroic materials.
On the other hand, both magnetic and ferroelectric orders arise from the same V cation, and there is even a parasitic polarization from $d$-electron, which presents a feature intrinsic to the type-II multiferroics.
Such a duality makes it highly possible to combine large polarization with strong magneto-electric coupling.

\begin{figure}[htb!]
 \centering
 \includegraphics[width=0.99\columnwidth]{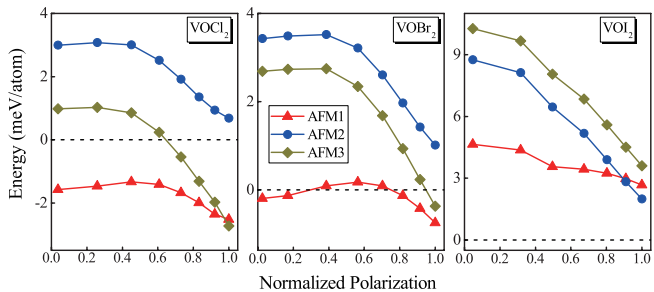}
 \caption{\label{Figure4}(Color online) Relative energy dependence on the polarization in VO$X_2$ for four spin configurations considered in Fig. 2(a). The energy of the FM configuration is taken as the energy reference zero at a particular polarization. During the calculations, we fix all the structural degrees of freedom as the ground state except for moving the V manually along the V-O chain. Normalized polarization 1 corresponds to the fully distorted structures of Fig. 1.}
\end{figure}

Finally, we tentatively evaluate the magneto-electric coupling by monitoring the magnetic response to the variation of ferroelectric polarization. We move the V atom away from the ground state along the V-O chain to change the polarization monotonously. Figure 5 shows the corresponding energy dependence of the four magnetic configurations on the polarization with ferromagnetic phase as the energy reference. Although the ground state is always ferromagnetic for VOI$_2$, we interestingly observe a crossover of the long-range magnetic order in VOCl$_2$ and VOBr$_2$ during the change of polarizations. That is, the long-range magnetic order changes from AFM3 to AFM1 for VOCl$_2$ while from AFM1 to FM for VOBr$_2$ with the decrease of the polarization. Besides, the energy differences between the different magnetic states vary remarkably with the polarization, reflecting the influence of polarization on the magnetic property. Such behaviors may be clues of sizable magneto-electric coupling.

In summary, we have predicted three monolayer multiferroic materials in violation of the $d^0$ rule. VOI$_2$ shows the fascinating ferromagnetism and ferroelectricity while VOCl$_2$ and VOBr$_2$ show the antiferromagnetism and ferroelectricity.
We unravel a novel mechanism for multiferroics beyond the $d^0$ rule, i.e., the filled $d$-orbital of V is perpendicular to the \emph{a priori} polarization and contributes a parasitic polarization, so as to realize the coexistence of $d$-electron ferroelectricity and magnetism. Our works thus offer a new way on searching for and engineering practical multiferroic materials.

\begin{acknowledgments}

This work was supported by the Basic Science Center Project of National Natural Science Foundation of China (Grant No. 51788104), the Ministry of Science and Technology of China (Grant No. 2016YFA0301001), the National Natural Science Foundation of China (Grant No. 11674071, No. 21773002, No. 11704038, and No. 11874089), the Open Research Fund Program of the State Key Laboratory of Low-Dimensional Quantum Physics (Grant No. KF201712), the Beijing Advanced Innovation Center for Future Chip (ICFC), and the Beijing Institute of Technology Research Fund Program for Young Scholars.

\end{acknowledgments}

$Note$ $added$: After we submitted the manuscript, we became aware of the work \cite{Nanoscale} which predicted the coexistence of ferroelectricity and antiferromagnetism in monolayer VOCl$_2$ and VOBr$_2$.


\end{document}